\documentclass[conference]{IEEEtran}
\IEEEoverridecommandlockouts
\usepackage{graphicx}
\usepackage{amssymb}
\usepackage{pifont}
\usepackage{multicol}
\usepackage{epsfig}
\usepackage{url}
\usepackage{hyperref}
\usepackage{subfigure}
\usepackage{multirow}

\usepackage{listings}
\usepackage{color}
\usepackage{graphicx}
\graphicspath{{./figures/}}
\usepackage{subfigure}
\usepackage{comment}

\newcommand*{\email}[1]{%
    \normalsize\href{mailto:#1}{#1}\par
    }

\begin{document}

\title{Fault Attacks on Secure Embedded Software: \\
Threats, Design and Evaluation
\thanks{This is a preprint version of Yuce, B., Schaumont, P. \& Witteman, M. {\em Fault Attacks on Secure Embedded Software: Threats, Design, and Evaluation}. J Hardw Syst Secur 2, 111–130 (2018). \url{https://doi.org/10.1007/s41635-018-0038-1}}
}


\author{\IEEEauthorblockN{Bilgiday Yuce}
\IEEEauthorblockA{Virginia Tech\\
Blacksburg, VA\\
\email{bilgiday@vt.edu}}
\and \IEEEauthorblockN{Patrick Schaumont}
\IEEEauthorblockA{Virginia Tech\\
Blacksburg, VA\\
\email{schaum@vt.edu}}
\and \IEEEauthorblockN{Marc Witteman}
\IEEEauthorblockA{Riscure -- Security Lab\\
Delft, Netherlands\\
\email{witteman@riscure.com}}
}




\maketitle

\begin{abstract}
Embedded software is developed under the assumption that hardware execution is always correct.
Fault attacks break and exploit that assumption. Through the careful introduction of targeted faults,
an adversary modifies the control-flow or data-flow integrity of software. The modified
program execution is then analyzed and used as a source of information leakage,
or as a mechanism for privilege escalation. Due to the increasing complexity of modern
embedded systems, and due to the difficulty of guaranteeing correct hardware execution 
even under a weak adversary, fault attacks are a growing threat. For example, the assumption
\emph{that an adversary has to be close to the physical execution of software, in order to inject}
an exploitable fault into hardware, has repeatedly been shown to be incorrect. 
This article is a review on hardware-based fault attacks on software, with emphasis on the 
context of embedded systems. We present a detailed discussion of the anatomy of a fault 
attack, and we make a review of fault attack evaluation techniques. The paper emphasizes the perspective from the attacker, rather than the perspective of countermeasure development. However, we emphasize that improvements to countermeasures often build on insight into the attacks.
\end{abstract}

\bibliographystyle{IEEEtran}

%
%

\section{Introduction}
\label{sec:intro}

In this paper we consider the fault attack threat against secure embedded software. Software plays a 
crucial role in the functionality of embedded computers. For example, 
in a System on Chip, software provides flexibility and it provides the logical integration 
of specialized hardware components. {\em Secure} embedded software is 
any software that employs security mechanisms (e.g, confidentiality, integrity, authentication and access control mechanisms) to ensure the security of sensitive data and functionality. Secure embedded software 
is therefore not only limited to cryptographic software, but also covers access control
and permission-rights management.  

\begin{table*}[t]
  \begin{center}
    \caption{Embedded Software Targets for the Hardware Attacker}
    \label{tab:faultexample}
    \begin{tabular}{lll} 
      \textbf{Abstraction Level} & \textbf{Cause of Security Failure} & \textbf{Attacker}\\
      \hline
      \hline
      Input/Output             & Software Bugs                   & Input/Output Attacker \\
      \hline
      Application Level        & Lack of Memory Region Isolation      & Memory Attacker        \\
      \hline
      Instruction Level        & Opcode modification                 & Hardware Attacker      \\
      Micro-Architecture Level & Instruction execution is wrong &                     \\
      Circuit Level            & Timing, Threshold Levels are not met  &                     \\
      Environment              & Operating Conditions are abnormal       &                     \\
      \hline
    \end{tabular}
  \end{center}
\end{table*}

Embedded computers that run secure embedded software are all around us. A large portion of the information ecosystem consists of embedded connected computers
that participate in the physical control and the measurement of critical infrastructures and utilities, such as smart grid, automotive and industrial controls. In addition, 
information technology is pervasive in the immediate vicinity of people such as in
cell phones, activity trackers, medical devices, or biometric tokens. The data handled by these embedded computers is sensitive. Computing devices in critical 
infrastructure execute safety-critical commands and collect sensitive measurement 
data protected by cryptographic keys and authentication codes. Human-centric information 
systems work with private end-user data, passwords, PIN codes, biometric data, location history, and usage patterns. Furthermore, the firmware and configuration data of embedded computing systems may 
represent valuable intellectual property.

The data and the unauthorized access of these embedded devices is an obvious target 
for attackers. At high level, the purpose of an attacker is to obtain control 
over the execution of the embedded software, or to extract internal data 
values processed by the embedded software.
We identify three different attackers, distinguished by the abstraction level
they operate on. The {\em input-output} attacker manipulates the data inputs of an 
embedded software application to trigger internal buffer overflows or internal 
software bugs in the application. The {\em memory} attacker co-exists with
the embedded software application, for example as a malicious software task, 
and snoops the memory space in order to directly manipulate or observe a secure 
embedded application~\cite{lipp2018meltdown,kocher2018spectre}. Both of these attackers succeed because they break an implicit 
assumption made by the secure embedded software application. The input-output attacker 
exploits the assumption that there are no malformed inputs to the program. 
Using malformed inputs, the attacker exploits bugs in secure embedded software 
such as missing memory bounds checks. The memory attacker exploits the assumption 
that memory space is private to the secure embedded application. This privacy
gets lost when the architecture cannot provide memory region isolation to
the application \cite{piessens16}

The third type of attacker, the {\em hardware fault} attacker, is the focus of this paper. 
Like the previous two, the hardware fault attacker breaks an implicit assumption made by the
secure embedded software application. In this case, the assumption is that the 
embedded hardware guarantees correct execution of the software. 
Table \ref{tab:faultexample} illustrates that the correct execution of software builds 
on many interdependent assumptions at different levels of abstraction in the hardware. 
Yet, {\em any} of the abstraction levels is a potential target for the hardware fault attacker. 

\begin{itemize}

\item At the {\em instruction-level}, a programmer assumes that the opcodes executed by 
the embedded hardware (the microprocessor) are correct. A hardware fault attacker who can 
manipulate opcodes can change the meaning of a program.

\item At the {\em micro-architecture level}, a programmer assumes that correct
opcodes imply correct execution of the instruction. A hardware fault attacker who can manipulate the micro-architecture can break this assumption and still change the meaning of
a program.

\item At the {\em circuit level}, the correct execution of software requires that digital logic in the processor will operate with the proper timing, and using the proper voltage levels to capture digital-0 and digital-1. 
A hardware fault attacker who can influence circuit timing or 
logic threshold levels, can change the correctness of digital execution and
hence still change the meaning of a program.

\item At the lowest abstraction level, the correct execution of software requires that the physical environment
of a digital circuit has nominal operating conditions, that it is using the
proper temperature, the proper circuit voltage, and the proper electromagnetic
environment. A hardware fault attacker who can influence any of these parameters
can change the correctness of the architecture, and hence still change
the meaning of a program.

\end{itemize}

This paper is only concerned with the hardware {\em fault} attacker, that is, an attacker
who builds on the manipulation of execution correctness at the architecture level or below.
We do not ignore side-channel attacks that build on the observation of the physical
effects of computing. In fact, some of the recent fault attacks successfully combine ideas of side-channel attacks with fault analysis. However, for this paper, we consistently develop the viewpoint of the hardware fault attacker as a threat to secure embedded software. An interested reader may refer to the existing works~\cite{witteman2008,yuce2016fame,barenghi12,joye12book,galathy2017systematic} for the viewpoint of the fault-resistant system designers.

It may appear as if a hardware attacker must be physically close to the digital
architecture to break execution correctness (at any abstraction level). However, 
this is not always the case. A hardware fault attacker can be a physical entity (in hardware)
or a virtual entity (in software). In the latter case, the attacker is logically present as somebody who runs attack software in conjunction with a victim program.
Recent work has shown that execution correctness of 
the hardware can be manipulated by such a software attacker. 

This paper addresses the following questions in detail.
\begin{itemize}
\item What are the common techniques for fault injection in a digital architecture, and
how do faults appear as a result of fault injection?
\item How do faults propagate through the micro-architecture and across the
architecture-level into the secure embedded software? 
\item How does the attacker exploit these faults towards a fault attack?
\item What testing equipment can be used to study the fault injection,
propagation and exploitation of secure embedded software?
\end{itemize}

This paper is organized as follows. In the next Section, we develop a systematic threat model against embedded software from the viewpoint of fault attacks, and we break down a fault attack into smaller steps. In Section Three, we describe commonly used fault injection techniques. In Section Four, we investigate the impact of fault injection on microprocessor execution. In Section Five, we analyze how faults propagate from the micro-architecture into the embedded software functionality. Section Six describes commonly used fault exploitation techniques. Section Seven describes fault attack evaluation technologies, and certification of embedded software against fault attacks.
We then conclude the paper.

\section{Background}
\label{sec:background}

\subsection{Threat Model}
The aim of a fault attack is breaching the security of a software program by forcing a security-sensitive asset into unintended behavior. For this purpose, the adversary injects well-crafted, targeted hardware faults by deliberately altering the operating conditions of the microprocessor that runs the target software. Then the adversary exploits the effects of the faults on the target software's execution and breaks the security. Consequently, the target of exploitation is the software layer while the origin of vulnerability (i.e, faults) is the hardware layer. 

In a typical fault attack, the adversary is not capable of directly modifying or monitoring the internals of a chip, or changing the binary of a program. The adversary is able to alter the execution of a target program by controlling the physical operating conditions (e.g, timing, supply voltage, temperature) of the processor hardware executing the program. The adversary can also provide input to the target program, and observe the effects of abnormal operating conditions on the software execution through system output or a related side-channel such as power consumption, cache-activity-related timing, and performance counters. 

\subsection{Using Faults as a Hacking Tool}


\begin{figure}[t]
  \centering
  \includegraphics[width=1\columnwidth]{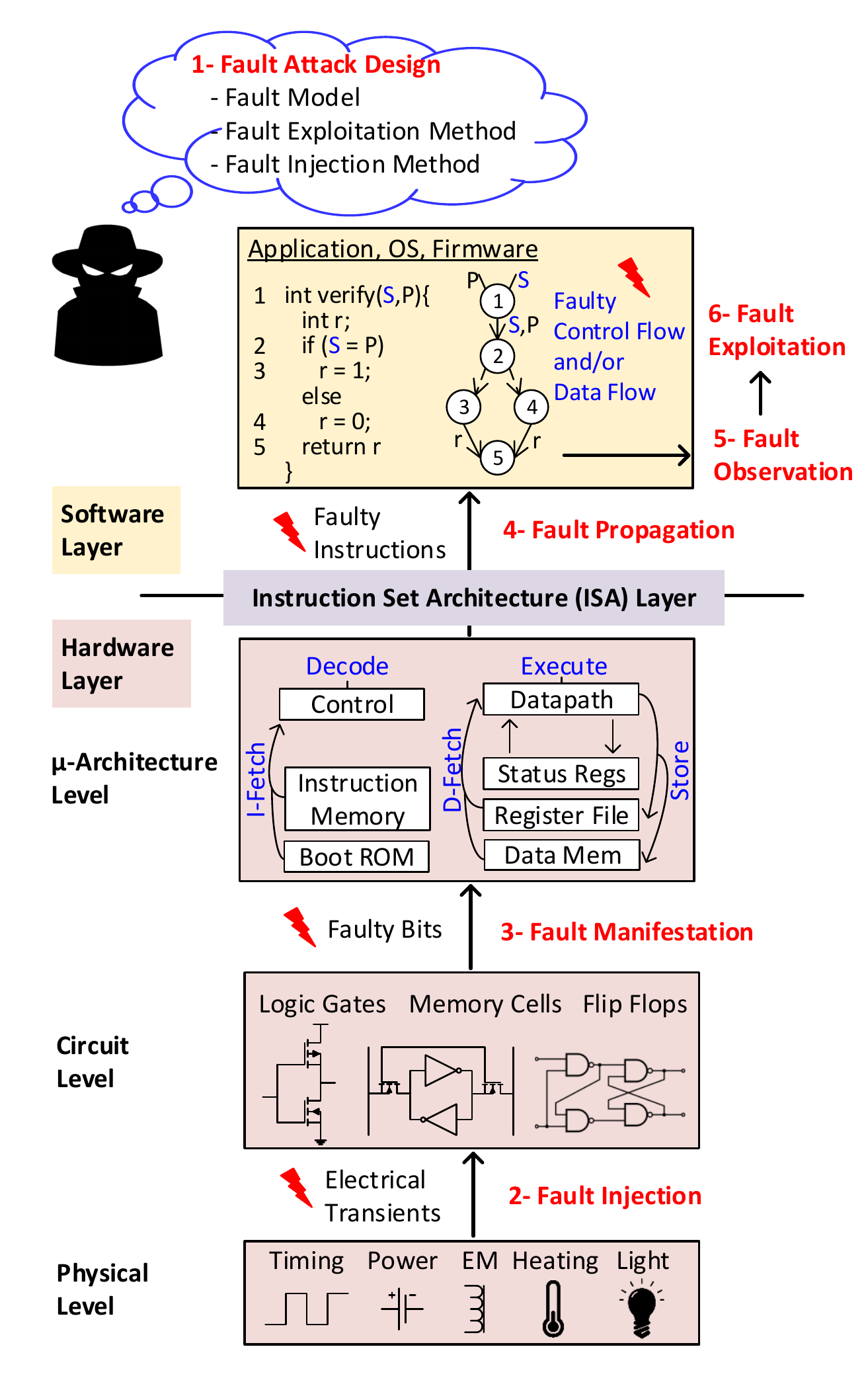}
  \caption{Anatomy of a typical fault attack on embedded software: The target of fault injection is the hardware while the target of exploitation is the software.}
  \label{fig:attack-overview}       
\end{figure}

Figure~\ref{fig:attack-overview} illustrates the steps and mechanisms involved in a typical fault attack on embedded software. A fault attack consists of two main phases, fault attack design and fault attack implementation (Steps 1-5 in Fig.~\ref{fig:attack-overview}). In the design step, the adversary analyzes the target to determine fault model (i.e, an assumption on the faults to be injected), fault exploitation method, and fault injection technique. For instance, an adversary may intend to inject faults into several assets such as an encryption program, a security-related verification code, a memory transfer function, the processor state register, a system call, the firmware, or configuration information of the target device. The adversary may then exploit the fault effects on the target asset for various attack objectives such as weakening the security, bypassing security checks, intellectual property theft, extracting the confidential data, privilege escalation, activating debug modes, and disabling secure boot of the device.

The implementation phase is a combination of five steps:
\begin{enumerate}
\item \textbf{Fault Injection:} In this step, the adversary applies a physical stress on the microprocessor to alter its physical operating conditions and to induce hardware faults. The applied physical stress can be in various forms such as clock glitches, supply voltage glitches, electromagnetic (EM) pulses, and laser shots.

To induce the desired faults, the adversary varies \textit{fault timing} and \textit{fault intensity}. Fault timing specifies when the physical stress is applied on the target processor. Fault intensity is the degree of the physical stress by which the microprocessor hardware is pushed beyond its nominal operating conditions. The adversary controls the fault intensity via fault injection parameters. For clock glitching, shortening the length of the glitch increases the fault intensity. It is controlled by glitch/pulse voltage and length for voltage glitching, electromagnetic pulse injection, and laser pulse injection. The laser and electromagnetic pulse injections also enable the adversary to localize the fault intensity by controlling the shape, size, and position of the injection probe.

\item \textbf{Fault Manifestation}: The circuit-level effect of fault injection is creating electrical transients on the nets, combinational gates, flip-flops, or memory cells. A fault manifests at the micro-architecture level when the electrical transients are captured into a memory cell or flip-flop, and change its value.

The number of manifested faulty bits in the micro-architecture level is correlated to the applied fault intensity: A gradual change in the fault intensity causes a gradual change in the manifested faults. We call this relation \textit{biased fault behavior}. This behavior is valid independent of the used fault injection method, and it enables the adversary to control the size (e.g, single-byte) of the induced faults~\cite{moro2013electromagnetic,courbon2014adjusting,yuce2015improving}. However, tuning the fault intensity alone is not sufficient to control the type (e.g, bit-set) and location (e.g, decode logic) of the induced faults. The adversary's control on the fault type and location is also affected by the type and precision of the fault injection equipment.  

The biased fault behavior also allows the adversary to find a critical fault intensity value, at which the electrical transients become strong enough to cause fault manifestation. That critical fault intensity value is called \textit{fault sensitivity} of the target hardware~\cite{li2010fault}.

\item \textbf{Fault Propagation}: In this step, the effects of the manifested faults are propagated to the software layer through execution of faulty instructions. The next two paragraphs briefly explain the mechanism behind fault propagation.

Software security mechanisms are implemented as a sequence of instructions executed by the microprocessor hardware. In addition, each instruction goes through the \textit{instruction-execution cycle} that consists of multiple steps carried out by a certain subset of available micro-architecture-level hardware blocks. The processor loads each instruction from program memory (\textit{instruction-fetch}), then determines the meaning of the current instruction through its opcode (\textit{instruction-decode}), then executes the current instruction (\textit{instruction-execution}), and then updates the state of the processor based on the instruction's result (\textit{instruction-store}). The number of steps in the instruction-execution cycle is architecture dependent, and it can vary considerably from one microprocessor to the next.

The manifested faults may cause faulty bits in any micro-architectural hardware block such as instruction memory, controller, datapath and register file. The effects of the manifested faults are propagated to the software layer when an instruction uses the affected micro-architectural block. As each instruction uses a specific subset of the micro-architectural blocks, the precise effect of a hardware fault depends on the type of the instruction. For instance, a bit-flip fault injected during the execution step of an addition instruction may yield a single-bit fault in the result of this instruction. However, the same bit-flip fault injected during a memory-load instruction would cause a single-bit fault in the effective address calculation, and thus, data is loaded from a wrong memory location. In the former case, only a single bit of the destination register is faulty; while in the latter case the destination register has a random number of faulty bits.

\item \textbf{Fault Observation:} An adversary needs to observe the effects faulty instructions in order to exploit them. An observable fault effect can be a faulty system output such as a faulty ciphertext, a side-channel information such as a sudden change in the power consumption, a single-bit information showing if fault injection was successful, or micro-architectural effects observed through performance counters~\cite{joye12book,bhattacharya2017formal}. These effects become observable to the adversary when they are subsequent instructions that have data-dependencies or control-dependencies on the faulty instruction are executed.

\item \textbf{Fault Exploitation:} In the final step, the adversary exploits the observable fault effects and breaks the security. For example, the adversary can analyze the differential of the correct and faulty ciphertexts from a cipher to retrieve the secret key used for the encryption. For the same purpose, an adversary may also use a single-bit side-channel information of whether fault injection was successful. Similarly, the adversary may use the faults to trigger traditional logical attacks such as buffer overflows and privilege escalation.

\end{enumerate} 

\section{Fault Injection Techniques}
\label{sec:fault-injection}
In a fault attack, it is essential to induce well-controlled faults during execution of the target software. An adversary achieves fault injection by deliberately applying physical stress to push the operating conditions of the underlying microprocessor hardware beyond their allowed margins. The adversary controls the induced faults through timing, location, and intensity of fault injection.  The {\em timing of fault injection} is defined as the moment at which physical stress is applied to the processor. The {\em location of the fault injection} is the spatial portion of the processor that is exposed to physical stress. The {\em intensity of the fault injection} is the amount of physical stress applied to the processor.

\begin{figure}[t]
	\centering
	\includegraphics[width=1\columnwidth]{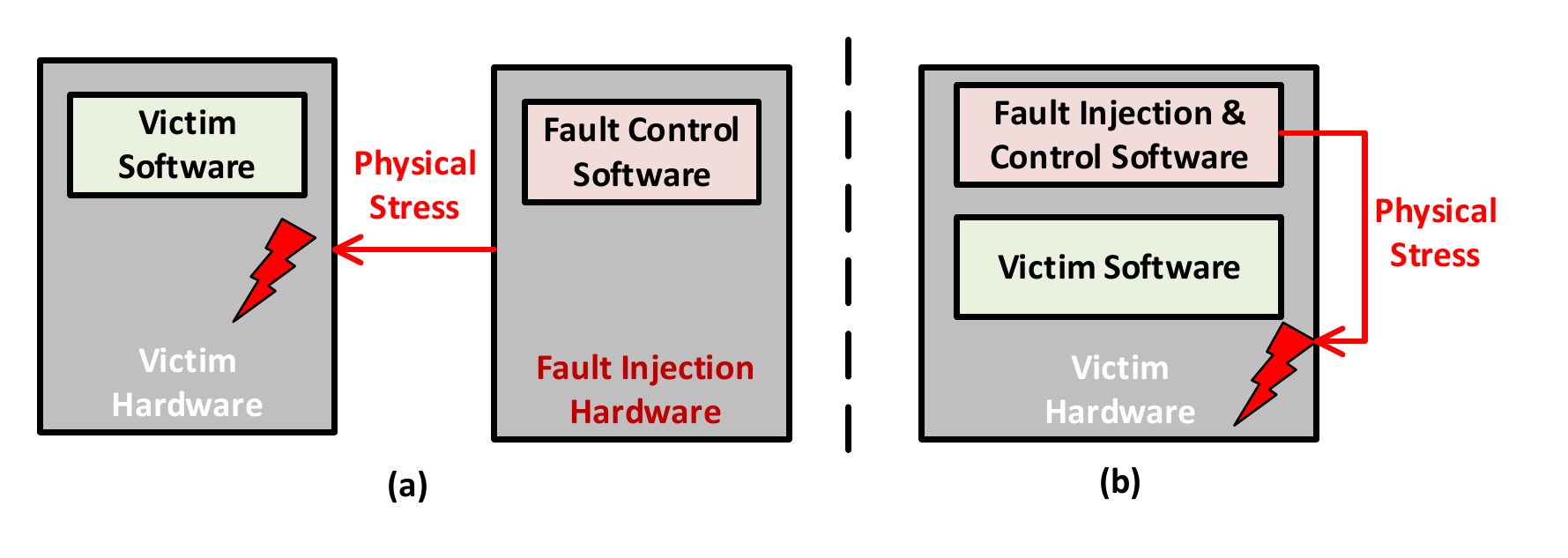}
	\caption{Fault Injection Categories: (a) Hardware-Controlled Fault Injection (b) Software-Controlled Fault Injection}
	\label{fig:injection-overview}       
\end{figure}

This section discusses common techniques used for fault injection. We briefly describe main characteristics of each fault injection technique. We partition the fault injection techniques into two main categories (Figure~\ref{fig:injection-overview}): Hardware-controlled fault injection and software-controlled fault injection. 

\textit{Hardware-controlled fault injection techniques} employ a separate external fault injection hardware to apply physical stress to the target hardware and induce faults in the victim software. Typically, the fault injection process is controlled by another software program (i.e, fault control software) running on the fault injection hardware. \textit{In software controlled fault injection techniques}, fault injection is controlled with a malicious software (i.e, fault injection and control software), which runs on the same hardware platform as the target software does. This malicious software alters the physical operating conditions of the target hardware to induce faults. While hardware-controlled techniques require physical proximity to the target system, software-controlled fault injection techniques enable remote fault attacks. 

\subsection{Hardware-controlled Fault Injection Techniques}
Several hardware-controlled fault injection techniques have been successfully demonstrated in the literature~\cite{bar2006sorcerer,joye12book}. The following sections provide an example list of commonly used techniques.

\subsubsection{Tampering with Clock Pin}

An adversary may inject faults by tampering with the external clock signal of the target device. 

One way of exploiting the clock signal for fault injection is {\em overclocking}~\cite{guilley2008silicon}, in which the adversary persistently applies a higher-frequency clock signal than the nominal clock frequency of the device. This violates setup-time constraints of the device and causes premature latching of the faulty values in flip-flops of the device~\cite{zussa2012investigation}. The spatial precision of this method is low because the modifications in the external clock signal are distributed across the whole chip surface through a clock network. Similarly, the temporal precision of overclocking is also low because all of the clock cycles are affected by fault injection; the adversary cannot select the clock cycles to be affected by the fault injection. On the other hand, the adversary has a fine control on the fault intensity through clock frequency. 


Another way of tampering with the clock signal is {\em clock glitching}~\cite{korakglitchpower2014}, in which the adversary temporarily shortens the length of a single clock cycle. This causes setup-time violations during the affected clock cycle. In comparison to overclocking, the adversary has a precise control on the temporal location (i.e, timing) of the fault injection. The intensity of the fault injection is controlled through the length of the glitched clock cycle. Similar to the overclocking, the spatial precision of clock glitching is low. 

For the clock glitching and overclocking techniques, the state-of-the-art fault injection setups~\cite{riscure,oflynn2014chipwhisperer} provide nanosecond-level temporal precision. The disadvantage of tampering with the clock signal is that this method requires physical access to an external clock pin. If a device uses an internally-generated clock signal, using this method is infeasible. 

\subsubsection{Tampering with Supply Voltage Pin}
An adversary can also inject faults by altering the external supply voltage of the target device. The adversary may use {\em underfeeding}~\cite{barenghi2009low}, in which a lower voltage than the nominal voltage is supplied to the device. Lower supply voltage increases the delay of combinational paths. This causes setup-time violation when the voltage drop is large enough to make a path delay larger than the applied clock period. This method has low spatial precision as the supply voltage is distributed all over the chip through a power network. Similarly, the temporal precision of the fault injection is low because all of the clock cycles are exposed to the lower supply voltage. The adversary controls the fault intensity through the value of the external supply voltage. 

The adversary can also use {\em voltage glitching}~\cite{timmers16}, which injects temporary voltage drops and provides the capability to control the temporal location of the fault injection. In this case, the adversary controls the intensity with the glitch offset from the sampling edge of the clock signal, glitch voltage, and glitch width similar to the clock glitching.   

These methods require physical access to the supply voltage pin. Removing the external coupling capacitance on the supply voltage line improves the efficiency~\cite{timmers16}. 
The drawback of tampering with external voltage pin is that the adversary does not have precise control on the timing and location of the fault injection.  

\subsubsection{Tampering with Operating Temperature}
An adversary may also use {\em overheating} to trigger setup-time violations~\cite{zussa2012investigation,hutter2013temperature} for fault injection. In this method, the adversary does not have precise control on the spatial and temporal location of the fault injection. The intensity of fault injection is controlled via operating temperature of the target device. 

In addition to the setup-time violation on the datapath, overheating also causes modification in memory cells in EEPROM~\cite{skorobogatov2009local}, Flash~\cite{skorobogatov2009local}, and DRAM~\cite{govindavajhala2003using} memories. While Govindavajhala et al.~\cite{govindavajhala2003using} use a low-spatial-precision light bulb as the heating source, Skorobogatov~\cite{skorobogatov2009local} employs a 650nm-wavelength laser to increase the spatial precision of heating. 

\subsubsection{Combination of Voltage, Frequency, and Temperature Fault Injection}
Zussa et al.~\cite{zussa2012investigation} demonstrated that overclocking, clock glitching, voltage glitching, underfeeding, and overheating exploit the same fault injection mechanism, which is the violation of a device's setup-time constraints. In addition, Korak et al.~\cite{korakglitchpower2014,korak2014clock} showed overheating and voltage glitching improves the efficiency of clock glitching.

\subsubsection{Optical Fault Injection}
In optical fault injection, the adversary decapsulates the target integrated circuit (IC) and exposes the silicon die to a light pulse. The applied light pulse induces a photo-electric current in the exposed area of the IC, which then cause faulty computations~\cite{skorobogatov2002optical}. The spatial location is controlled by the position and the size of the light source, and the temporal location is controlled by the offset of the pulse from a trigger signal. The intensity of the fault injection is determined by the energy and duration of the light pulse. It has been demonstrated that optical fault injection can be achieved with a low-cost camera flash light~\cite{schmidt2007optical,skorobogatov2002optical}. The state-of-the-art optical fault injection setups~\cite{van2011practical} use laser beams for fault injection to achieve micrometer-level spatial and nanosecond-level temporal precision. They also provide precise control on the fault intensity. This enables an adversary to target a single transistor. Laser fault injection can be done from front side and back side of an IC. Front side attacks typically use light with shorter wavelengths. These beams have more energy, and can easily penetrate between metal layers. Back side attacks use infrared light that penetrates the silicon substrate without being blocked by metal layers. 

A disadvantage of the optical fault injection is that it requires decapsulation of the target IC. In addition, it can permanently damage the target IC. Despite these disadvantages laser fault injection is popular because it provides the most precise and effective fault injection means.



\subsubsection{Electromagnetic Fault Injection}
In electromagnetic fault injection (EMFI), the adversary applies transient or harmonic EM pulses on the target integrated circuit (IC) through a fault injection probe, which is designed as an electromagnetic coil. The adversary places the probe above the target IC and applies a voltage pulse to the coil, which induces eddy currents inside the target IC. Then the effects of the induced eddy currents are captured as faults. The adversary controls the temporal location of fault injection through offset of the EM pulse from a trigger signal. The spatial location of the fault injection is controlled via position and size of the injection probe. The fault intensity is determined by the voltage and duration of the applied EM pulse. The feasibility of EMFI on off-the-shelf microprocessor ICs has been demonstrated using both low-cost and high-cost injection setups. For instance, Schmidt et al.~\cite{schmidt2007optical} use a simple gas lighter to induce EM pulses onto an 8-bit microcontroller with low spatial and temporal precision. The state-of-the-art EMFI setups~\cite{maistri2014electromagnetic,moroemmodel14,velegalati2013electro} provide millimeter-level precision in spatial location and nanosecond-level precision in the temporal location of the EM pulse. Furthermore, these setups also provide precise control on the voltage and duration of the applied EM pulse. The advantages of EMFI is that it does not require decapsulation of the target IC and it can inject local faults. However, its spatial precision is lower than the spatial precision of the laser fault injection. 


\subsection{Software-controlled Fault Injection Techniques}
Software-controlled fault injection is a recently discovered research area. The following two sections briefly explain the existing two software-controlled fault injection techniques.

\subsubsection{Tampering with DVFS Interface}
In the modern embedded systems, Dynamic Voltage Frequency Scaling is a commonly used energy management technique, which regulates the operating voltage and frequency of a microprocessor based on its dynamic workload. In a typical DVFS scheme, kernel-level drivers control the frequency and voltage of a processor through on-chip regulators. 

Tang et al.~\cite{tang2017clkscrew} demonstrated that an adversary can exploit the interface between the software drivers and hardware regulators to induce faults in a multi-core processor. In this technique, the adversary uses a malicious kernel-level driver running on a processor core to set the operating voltage and frequency of another core that executes the victim software. This method allows an adversary to violate setup time constraints of the victim core via overclocking and underfeeding it for a specific period of time. The adversary controls the temporal location with the endpoints of the overclocking or underfeeding period. As both the clock and voltage signals are chip-level global signals, the adversary does not have a direct control on the spatial location. The intensity of fault is determined by the overclocking frequency and the underfeeding voltage value. This method requires neither additional fault injection hardware nor physical access to the target device.

\begin{table*}
	\centering
	\caption{Fault Injection Techniques and the Characteristics of the Corresponding Physical Stress Applied to a Target Device}
	\label{table:fi-summary}
	\resizebox{\textwidth}{!}{\begin{tabular}[t]{|l|l|l|l|l|}
		\hline
		\textbf{Fault Injection Technique}				& \multicolumn{4}{|c|}{\textbf{Characteristics of the Applied Physical Stress}}  \\ \hline
        				& Spatial Precision  & Temporal Precision & Cost & Controlling the Intensity \\
		\hline
		\hline
		Overclocking    & Low (Global)    & Low (Global) & Low & Clock Frequency \\ \hline
		Clock Glitching & Low (Global)    & High (Local) & Low & Glitch Width \\ \hline
        Underfeeding & Low (Global)    & Low (Global) & Low & Voltage Level \\ \hline
        Voltage Glitching & Low (Global)    & High (Local) & Low & Glitch Voltage \\
         &     &  &  & Glitch Width \\ \hline
        Overheating & Low (Global)    & Low (Global) & Low & Ambient Temperature \\ \hline
        Light Pulse & Medium (Local)    & Medium (Local) & Low & Pulse Width \\ 
        &     &  &  & Pulse Energy \\ 
         &     &  &  & Pulse Offset \\ \hline
        Laser Pulse & High (Local)    & High (Local) & High & Pulse Width \\ \cline{1-2} 
        EM Pulse &  Medium (Local)   &  &  & Pulse Energy \\ 
         &     &  &  & Pulse Offset \\
         &     &  &  & Probe Size \\ \hline
        DVFS interface & Low (Global)    & Medium (Local) & Zero & Supply Voltage \\ 
        &     &  &  & Clock Frequency \\ \hline 
        Memory Disturbance & High (Local)    & Medium (Local) & Zero & Disturbance Frequency \\  \hline
	\end{tabular}}
\end{table*}

\subsubsection{Triggering Memory Disturbance Errors}

In this fault injection method, the adversary injects faults into memory cells by exploiting the reliability issues of modern memory hardware such as DRAM and Flash memory chips. The continuous scaling down in the process technology has enabled memory manufacturers to significantly reduce cost-per-bit by placing smaller memory cells closer to each other. However, this also increases electrical interference between memory cells: Accessing a memory cell electrically disturbs nearby memory cells. A disturbed memory cell loses its value and experiences a memory disturbance error when the amount of electrical disturbance is beyond noise margins of that disturbed cell~\cite{cai2017vulnerabilities,kim2014flipping}.

An adversary may trigger memory disturbance errors through a non-privileged fault injection program. This program repeatedly accesses a set of memory cells (i.e, aggressor memory cells) to induce disturbance errors in a set of victim memory cells storing security-sensitive data. This method allows an adversary to corrupt memory space of a security-sensitive program from memory space of the adversary-controlled fault injector program. Memory disturbance errors have been demonstrated on commodity DRAM and NAND Flash memory chips~\cite{cai2017vulnerabilities}.


In DRAM memories, the memory disturbance errors are induced through Rowhammer mechanism~\cite{kim2014flipping}. Thus, it is called Rowhammering. A DRAM memory is internally organized as a two-dimensional array of DRAM cells, where each cell consists of an access transistor and a capacitor storing charge to represent a binary value. As capacitors lose their charges because of the leakage current, the DRAM cells are periodically refreshed to restore their charges. Each row of the array has a separate wordline, which is a wire connecting all memory cells on the corresponding row. To access a DRAM cell within the two-dimensional array, the corresponding row of the array is activated by raising the voltage of its wordline. Persistent access to the same row causes repeated voltage fluctuations on its wordline, which electrically disturbs nearby rows. This disturbance increases the charge leakage rate in the nearby DRAM rows~\cite{kim2014flipping}. As a result, a memory cell within a nearby row experiences a memory disturbance error (a bit-flip error) if it loses a significant amount of charge before it is refreshed. An adversary may take advantage of that physical phenomenon to inject faults. For this purpose, the adversary runs a malicious fault injection program on the target processor, which aims at altering a security-sensitive state of a victim program running on the same processor. The fault injection program continuously accesses an aggressor DRAM row in its own memory space and induces faults into a victim DRAM row within the victim program's memory space~\cite{gruss2016rowhammer,van2016drammer,razavi2016flip}.       

Similar disturbance mechanisms have been also demonstrated on multi-level cell (MLC) NAND Flash memories. Similar to DRAM memory, a Flash memory is also internally organized as an array of Flash memory cells, each of which is a floating-gate transistor. The amount of charge stored in the floating gate determines the threshold voltage of the transistor, which is used to represent the stored data, In MLC Flash memories, each cell stores two bits of data. Unlike DRAM memories, Flash memories do not require periodic refreshing. Cai et al.~\cite{cai2017vulnerabilities} demonstrated that the capacitive coupling between neighboring Flash cells enables two memory disturbance error mechanisms. The first mechanism, {Cell-to-Cell Program Interference (CCI)}, introduces faults into a Flash cell when a nearby cell is programmed (i.e, written). The amount of interference is high when a specific data pattern for programming is used. Cai et al.~\cite{cai2017vulnerabilities} and Kurmus et al.~\cite{kurmus17random} showed how a malicious fault injection program may trigger CCI mechanism to cause a security breach. The second mechanism is {\em Read-Disturb}, in which the content of a Flash cell is disturbed when a nearby cell is read. Cai et al.~\cite{cai2017vulnerabilities} demonstrated the use of read-disturb to cause security problems.

The advantage of fault injection by triggering memory disturbance errors is that it can induce single-bit to multi-bit faults into a certain memory location~\cite{razavi2016flip}. This enables an adversary to break several security mechanisms. 

Table~\ref{table:fi-summary} summarizes the previously described fault injection techniques and the main characteristics of the physical stress applied to the target device. For each fault injection technique, the table provides spatial precision, temporal precision, hardware cost of the applied physical stress. The table also provides a list of fault injection parameters to control the intensity of the applied physical stress.

\section{Fault Manifestation in Processor Micro-architecture}
\label{sec:fault-manifestation}

This section explains the effects of physical fault injection on the micro-architecture of the target processor. First, we will distinguish micro-architecture (i.e, internal architecture) of a processor from its architecture (i.e, external architecture). Then we will briefly explain main characteristics of the induced faults into micro-architecture.

Any processor can be described from two distinct architectural perspectives. The architecture of the processor describes it as seen by programmers in terms of its instruction set and facilities. The architecture defines semantics and syntax of available instructions, program-visible processor registers, memory model, and how interrupts are handled. It is the boundary between hardware and software as well as a contract between programmers and hardware designers. The micro\--ar\-chi\-tec\-ture describes the physical organization and implementation of the architecture. This includes the memory hierarchy, pipeline structure, available functional units, employed mechanisms (e.g, out-of-order execution) for instruction-level parallelism, and so forth. The micro-architecture is optimized to satisfy cost and performance requirements.

Faults manifest as incorrect bits in the flip-flops or memory cells employed in the micro-architecture if the applied physical stress is beyond noise margins of target processor hardware. The parameters and type of physical fault injection technique determine characteristics of the manifested faults. In this work, we use four parameters to describe any manifested fault in micro-architecture level:
\begin{itemize}
	\item \textbf{Location of the Manifested Fault:} This parameter specifies the micro-architectural blocks that contain faulty bits because of physical fault injection.
    
Faults may manifest in any micro-architectural block in the control or datapath part of the processor such as instruction memory, instruction fetch block, instruction decode block, operand fetch block, execution block, data memory, register file, processor status register, and conditional flags. An adversary's control on the location of the manifested faults depends on the spatial precision of the used fault injection method, which is characterized as precise control, loose control, and no control~\cite{karaklajic2013}. 

All of the previously described fault injection techniques, except software-based memory disturbance, can target the datapath, control, or memory of a processor. Memory disturbance can only inject faults into the memory.
	
	\item \textbf{Size of the Manifested Fault:} This parameter specifies the number of faulty micro-architecture bits induced by physical fault injection. An adversary can control the size of the manifested faults by adjusting the fault intensity. In the literature, manifested faults are commonly classified as single-bit faults, byte-size faults, word-size faults, and arbitrary-size faults~\cite{otto2005fault}. The adversary influences the size of the manifested faults by tuning the fault injection intensity.
	
	\item \textbf{Effect of the Manifested Fault:} This parameter specifies the logical effect of the manifested fault on the fault location. Common fault effects are stuck-at fault, bit-flip fault, bit-set fault, bit-reset fault, and random fault~\cite{otto2005fault}.
    
    All of the aforementioned fault injection techniques are able to induce bit-flip effects. In addition, laser and EM pulses are also able to induce bit-reset, bit-set, and stuck-at faults. 
	
	\item \textbf{Duration of the Manifested Fault:} Fault attacks typically exploit transient faults, which last as long as the physical stress is applied. Such faults are recovered when a new value is written into the faulty flip-flop or the memory cell. However, if a register is refreshed only infrequently, these faults can still last a long time. Some fault attack injection techniques are able to create long-lasting faults, such as recently demonstrated using focused X-ray injection \cite{anceau17}. Some fault injection techniques can even inject permanent faults, such as when a laser pulse causes permanent damage to a memory or register cell (stuck-at).
	
\end{itemize} 

The next section explains how the manifested faults propagate to the software layer.


\section{Fault Propagation to the Software Layer}
\label{sec:faultpropagation}

The manifested faults propagate to the software layer as faulty instructions when the micro-architectural blocks containing the faulty bits are used by the instructions of the target program. Propagated fault effects are determined by the type of affected instruction, type of faulty micro-architectural block, and the characteristics (size, effect) of the manifested faults. As each processor implementation has its own micro-architecture, it is not possible to list all of the potential fault effects propagated to software. Instead, we provide an example to demonstrate a list of potential fault effects for a subset of SPARC instructions running on a hypothetical generic micro-architecture. Using the same approach, similar lists can be built for specific instruction sets and processor implementations.

As an example, we chose four SPARCv8 instructions: a memory-load ({\tt ld}), a logic ({\tt xor}), a comparison ({\tt cmp}), and a conditional branch ({\tt be}) instruction. Table~\ref{table:sparc-instr} lists the instructions and their definitions. 

\begin{table}
	\centering
	\caption{An Example Set of SPARCv8 Instructions}
	\label{table:sparc-instr}
	\resizebox{\columnwidth}{!}{\begin{tabular}[t]{ll}
		\hline
		\textbf{Instruction}				& \textbf{Definition}  \\
		\hline
		\hline
		{\tt ld [r1+r2], r3}    & Loads a 32-bit word into register {\tt r3}    \\
		& from data memory (D-Mem) address {\tt r1+r2}.     \\ \hline
		{\tt xor r1, r2, r3}    & Bit-wise XOR operation on {\tt r1} and {\tt r2}    \\
		& Result is written to register {\tt r3}.     \\ \hline
		{\tt cmp r1, r2}    	& Compares registers {\tt r1} and {\tt r2} and    \\
		& updates conditional flags accordingly.     \\ \hline
		{\tt be offset}      		& PC-relative conditional jump:      \\ 
		& If zero-flag is set, PC will be {\tt PC + offset}. \\
        & Otherwise, PC will be {\tt PC + 4}. \\ \hline
	\end{tabular}}
\end{table}

\begin{table*}
	\centering
	\caption{Propagated effects to software layer for each faulty micro-architectural block (with 1-bit fault) and instruction}
	\label{table:fault-prop}
	\resizebox{\textwidth}{!}{\begin{tabular}[t]{|l||l|l|l|l|}
		\hline
		\textbf{Faulty Block}				& \multicolumn{4}{|c|}{\textbf{Propagated Fault Effects}}  \\ \hline
		(1-bit Fault)			& 	{\tt ld [r1+r2], r3} & {\tt xor r1, r2, r3} & {\tt cmp r1, r2} & {\tt be dest} \\ 
		\hline
		\hline 
		I-Mem	& 	 \multicolumn{4}{|l|}{Execution of a wrong instruction due to opcode-field corruption} \\
		I-Fetch (PC, IR)		& 	 \multicolumn{4}{|l|}{Fetching operands from wrong location due to source-operand-location corruption} \\
		I-Decode	& 	 \multicolumn{4}{|l|}{Updating a wrong destination due to destination-operand-location corruption} \\
					& 	 \multicolumn{4}{|l|}{Fetching next instruction from a wrong address due to PC corruption} \\ \hline
		O-Fetch		& 	 Arbitrary \# of faults in {\tt r3}  & 1-bit fault in {\tt r3} & Faulty update  & No \\
		(1-bit fault in {\tt r1} or {\tt r2})			& (Faulty D-Memory address)   & (Faulty XOR input(s)) & of conditional flags & effect \\ \hline
		Execute		& 	Arbitrary \# of faults in {\tt r3}  & 1-bit fault in {\tt r3} & Faulty update  &  1-bit fault in jump address \\
					& (Faulty D-Memory address)   & (Faulty XOR operation) & of conditional flags &  or Inversion of branch \\ \hline 
		Store		& 	1-bit fault in {\tt r3}  & 1-bit fault in {\tt r3} & Faulty update  & 1-bit fault in jump address  \\
		& (Faulty update of {\tt [r1+r2]})   & (Faulty update of {\tt r3}) & of conditional flags&  \\ \hline 
		D-Mem & \multicolumn{4}{|c|}{No effect} \\ \hline
		Register File & \multicolumn{3}{|l|}{Fetching wrong source operands from register file} & No effect\\ \hline
		Conditional Flags & \multicolumn{3}{|c|}{No effect} & No jump to {\tt dest} \\ \hline
	\end{tabular}}
\end{table*}  

The assumed generic micro-architecture contains the following blocks to carry out instruction-execution cycle for each instruction:

\begin{itemize}
	\item \textbf{I-Mem Block} is the instruction memory that stores the instructions.
	
	\item \textbf{I-Fetch Block} prepares the address for the instruction memory, program counter (PC). Then, using the prepared PC, it fetches an instruction into the instruction register (IR).
	
	\item \textbf{I-Decode Block} takes the fetched instruction from IR and decodes it to determine the location of the source operands, the location of destination operands, and the operation to be applied. The source operands are fetched from the register file. The destination may be the register file, data memory (D-Mem), or conditional flags.
	
	\item \textbf{O-Fetch Block} uses the decoded information to fetch the input operands from the register file and to feed them to the execution block. The {\tt be} instruction does not use this block because it does not fetch any operand from the register file.
	
	\item \textbf{Execute Block} applies the required operation on the fetched source operands and generates a result. For {\tt ld}, it calculates the D-Mem address from {\tt r1} and {\tt r2}. For {\tt xor}, it applies bitwise XOR operation on {\tt r1} and {\tt r2}. For {\tt cmp}, it subtracts {\tt r2} from {\tt r1}.  For {\tt be}, it calculates the destination address from the current PC and {\tt offset}. It also checks the conditional flags to determine if the branch will be taken.   
	
	\item \textbf{Store Block} updates the destination location (D-Mem, register file, or flags) with the result computed by the execution block. For the {\tt ld} and {\tt xor}, it is the register {\tt r3}. For the {\tt cmp} instruction, the destination is conditional flags. For the {\tt be} instruction, the destination is the PC value if the branch is taken. Otherwise, it will not affect any destination.  
\end{itemize}

Table~\ref{table:fault-prop} provides an example list of propagated fault effects for each (\textit{instruction, micro-architecture block}) pair. In this example, we assume a \textit{single bit-flip fault} in any micro-architectural block. A fault induced in I-Memory, I-Fetch, or I-Decode block would affect syntax (i.e, opcode and operands) and/or semantics (i.e, the operation to be applied) of an instruction independent from the type of the instruction. Thus, Table~\ref{table:fault-prop} shows the propagated fault effects for these blocks in a single cell. Faults induced in the other blocks would cause errors in the instruction-specific computation of a correctly fetched and decoded instruction:

\begin{itemize}
	\item \textbf{I-Mem, I-Fetch, I-Decode:} If the fault manifests in the opcode part of the faulty instruction, another instruction will be executed. If the fault affects the addresses of source operands, they will be fetched from an incorrect location. Similarly, the result of an instruction will be written into a wrong location if the fault hits the destination address. Finally, the next instruction will be fetched from an incorrect location if the PC calculation gets faulty.
	
	\item \textbf{O-Fetch:} For the {\tt ld} instruction, a single-bit fault in this block affects the value of register {\tt r1} or {\tt r2} fetched from register file. The fault then causes D-Mem address to be faulty. As a result, a single-bit fault in either {\tt r1} or {\tt r2} may induce an arbitrary number of faults in the destination register {\tt r3} because the result will be fetched from an incorrect D-Mem location.
    
    For the {\tt xor} instruction, the fault will affect a single bit of {\tt r1} or {\tt r2}, which will be propagated to {\tt r3} as a single-bit fault.
    
    For the {\tt cmp} instruction, the single-bit fault may affect the result of the comparison, which will alter the conditional flags based on the modified comparison result.
    
    For the {\tt be} instruction, the fault will not have any effect because this instruction does not fetch anything from the register file.
	
	\item \textbf{Execute:} For the {\tt ld}, {\tt xor}, and {\tt cmp} instructions, the effects of the fault will be same as the effects described in the O-Fetch case.
	
	For the {\tt be} instruction, the fault will change the single-bit of the computed branch address. If the branch is taken, the destination address will be wrong. Otherwise, the faulty branch address will not affect the program. The fault may also change the direction of the branch instruction from taken branch to non-taken branch, or vice versa.
	
	\item \textbf{Store:} For the {\tt ld} instruction, the fault will cause a single-bit error in the correctly computed D-Mem address {\tt [r1+r2]}. For, the {\tt xor}, and {\tt cmp} instructions, the effects of the fault will be same as the effects described in the O-Fetch case. For the {\tt be} instruction the fault will change the value of the PC if it is a taken branch.
	
	\item \textbf{D-Mem:} As none of the instructions use a value from the data memory, the fault in D-Mem will not affect any of the considered instructions.
	
	\item \textbf{Register File:} For the {\tt ld}, {\tt xor}, and {\tt cmp} instructions, the effects of the fault will be same as the effects described in the O-Fetch case. As the {\tt be} instruction does not use this block, the fault will not have any effect on this instruction.
	
	\item \textbf{Conditional Flags:} The fault in conditional flags will affect only the {\tt be} instruction as the other instructions do not use the conditional flags.
	   
\end{itemize}
\section{Fault Exploitation Techniques}
\label{sec:fault-exploitation}

This section presents main fault exploitation techniques, which have been proposed to break the security of both cryptographic and non-cryptographic security mechanisms protecting embedded software. Each exploitation technique relies on a fault model, which is a high-level assumption for the effects of physical fault injection on the execution of the target software. Thus, we start with commonly used fault models in practice. Then we will briefly explain fault exploitation techniques.

\subsection{Fault Models}  

In the design phase of a fault attack, an adversary makes a fault model assumption and develops an exploitation strategy based on the fault model. This assumption generally includes the location of the fault in the data or control flow of the target program, the timing of the fault with respect to the duration of the target program, size of the fault, and effect of the fault. The fault models can be described in algorithm level, source code level, or instruction level. The following paragraphs provide an example list of commonly used fault models.

The most of fault-based cryptanalysis techniques on symmetric and asymmetric cryptography assume faults on data flow of a target program that corrupt a single bit, single byte, multiple bytes, or a single word of a security-critical variable in various ways (e.g, flip, set, reset, random)~\cite{barenghi12,joye12book,otto2005fault}. 

On the control flow, the most popular fault models are to skip the execution of a specific instruction (i.e, instruction skip)~\cite{barbu2010attacks,dehbaoui2013electromagnetic}, multiple instruction skips~\cite{riviere2015high,nashimoto2017buffer}, replacing an instruction with another one (i.e, instruction modification)~\cite{balasch11,timmers16}, changing the result of a conditional branch~\cite{vetillard2010combined,potet2014lazart}, and tampering with loop counters~\cite{choukri2005round,dutertre2012fault}.

In the implementation of a fault attack, the adversary aims at inducing the fault effects assumed in the fault model via fault injection, fault manifestation, and fault propagation processes. Therefore, a fault model can be realized through different combinations of fault injection, fault manifestation, and fault propagation. The following sections provide a list of commonly used fault exploitation techniques to breach the security of embedded software.

\subsection{Cryptanalysis using Fault Injection}
Using fault injection for cryptanalysis has been extensively studied on the implementations of symmetric-key, public-key, and post-quantum cryptography algorithms~\cite{barenghi12,karaklajic2013,bar2006sorcerer,otto2005fault,joye12book}.

\paragraph{\textbf{Differential Fault Analysis (DFA)}} is the most widely used fault-based cryptanalysis technique. The main principle of DFA is to exploit the difference between the faulty and fault-free outputs of a cryptosystem. In a typical DFA attack, an adversary collects two outputs (e.g, ciphertexts) from a cryptosystem (e.g, encryption) that are generated for the same input (e.g, plaintext) and secret variable (e.g, encryption key). One of the outputs is collected without fault injection. During the generation of the second output, the adversary injects a certain fault into the execution of the cryptosystem. Then the adversary analyzes the propagation of this fault differential to the output and reveals the secret variable. DFA attacks assume specific fault during differential analysis of the faulty and fault-free outputs. Various DFA techniques have been successfully demonstrated on block ciphers~\cite{biham1997differential}, stream ciphers~\cite{hoch2004fault}, public-key algorithms~\cite{biehl2000differential,joye12book}, and post-quantum cryptography~\cite{taha2015implementation}. To illustrate how the DFA works, the following two paragraphs briefly explain a previously proposed DFA attack on Advanced Standard Encryption (AES) algorithm. 
\begin{figure}[t]
	\centering
	\includegraphics[width=0.6\columnwidth]{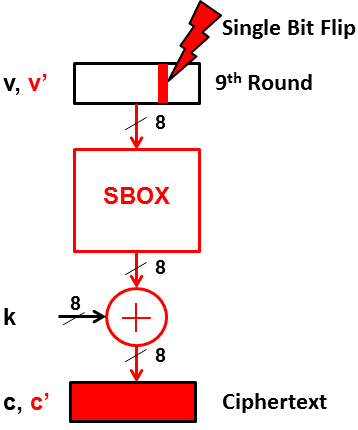}
	\caption{Propagation of a single bit flip fault through the last round of AES. The fault causes a single faulty byte in the ciphertext.}
	\label{fig:dfa-on-aes}       
\end{figure}

In this example attack~\cite{giraud2004dfa}, the assumed fault model is a single bit flip in the input of the last AES round. Figure~\ref{fig:dfa-on-aes} shows the propagation of a single bit fault through the last round of AES. Due to the structure of AES, the single bit flip (\textit{v*}) is propagated to the ciphertext as a single fault byte (\textit{c*}). For the fault-free (\textit{c}) and faulty (\textit{c*}) ciphertexts, we can write
\begin{lstlisting}
c  = SBOX(v) $\oplus$ k
c* = SBOX(v*) $\oplus$ k
$\Delta$  = v $\oplus$ v*
\end{lstlisting}
where $\Delta$ denotes the injected fault differential. The adversary is able to observe the values of \textit{c*} and \textit{c}. As the assumed fault model is to flip a single bit of \textit{v}, the Hamming Weight (HW) of the injected fault differential $\Delta$ is assumed to be 1. The purpose of the adversary is to reveal the value of the  corresponding byte ($k$) of the last round key. The adversary achieves this through an exhaustive search on the possible key values. For each possible key hypothesis $\tilde k$ for actual key byte $k$, the adversary first computes the corresponding fault differential $\tilde \Delta$ as follows: 
\begin{lstlisting}
$\tilde v$  = SBOX$^{-1}$(c $\oplus$ $\tilde k$)
$\tilde v$*  = SBOX$^{-1}$(c* $\oplus$ $\tilde k$)
$\tilde \Delta$ = $\tilde v$ $\oplus$ $\tilde v$*
\end{lstlisting}
The adversary then checks whether the computed (i.e, hypothesized) differential $\tilde \Delta$ is equal to the injected (i.e, assumed) differential $\Delta$. For our example, the hypothesized key $\tilde k$ is a possible candidate for the actual key $k$ if the Hamming weight of $\tilde \Delta$ is 1. Otherwise, the hypothesized key $\tilde k$ is discarded.

After testing all of the possible key hypotheses, the set of possible key candidates contains 8 elements on the average~\cite{ferretti2014role}. Therefore, two fault injection experiments on a given input byte of the AES last round reveals the value of the corresponding byte of the AES last round key. The remaining bytes of the last round key can be revealed by repeating the same steps for the remaining input bytes of the last round. As a result, the explained attack requires 32 fault injection experiments to retrieve the whole 16-byte last-round key of AES-128. The adversary can then calculate other round keys by applying AES key scheduling algorithm on the retrieved last round key. For further information on the DFA techniques and their comparison, the reader may refer to existing works in the literature~\cite{sakiyama2012information,ali2013differential,joye12book}. 

\paragraph{\textbf{Biased Fault Analysis}} attacks~\cite{ghalaty2014differential,li2012new,liu2015dera,fuhr2013fault,jarvinen2012harnessing} exploit biased fault behavior: Because of the correlation between the fault behavior of a target program and the applied physical fault intensity, the distribution of fault models is non-uniform. They allow an adversary to treat fault behavior as a side-channel signal, which relaxes the strict fault model requirements of the previous attacks~\cite{li2010fault}. As an example of biased fault analysis, we will demonstrate Differential Fault Intensity Analysis (DFIA) on AES, which was proposed by Ghalaty et al.~\cite{ghalaty2014differential}.

Similar to the previous DFA example, the DFIA assumes faults in the input of the AES last round. Unlike DFA attacks, DFIA does not make a precise assumption on the injected faults. Instead, DFIA assumes that the injected faults are biased: The adversary adjusts the fault intensity during the fault injection step such that the number of faulty bits in the input of the last AES round (\textit{v}) is minimal. For this specific DFIA attack, it is assumed that the number of faulty bits in the input byte \textit{v} is less than 4~\cite{ghalaty2014differential}. DFIA retrieves the value of the corresponding key byte as follows. The adversary first computes the fault differential  $\tilde \Delta$ for each key hypothesis  $\tilde k$. If the correct key hypothesis is made, the Hamming weight of the hypothesized fault differential $ \tilde \delta$ is small (i.e, $ HW(\tilde \delta) < 4$). Under a wrong key hypothesis, the expected Hamming weight of the fault differential is large because of highly non-linear design of SBOX of AES. 

It is possible that a single fault is insufficient to uniquely determine the correct key. In that case, the adversary can inject multiple biased faults, under a gradually increasing fault intensity, each time recording the faulty ciphertext \textit{c*}. For each key hypothesis $\tilde k$ and injected fault, the adversary computes the hamming weight of the corresponding fault differential. For the correct key hypothesis, the sum of all Hamming weights is still minimal~\cite{ghalaty2014differential}. Ghalaty et al. demonstrated that, on the average, DFIA requires 4.6 faults to retrieve one byte of the last round key and 68 faults to retrieve all bytes of it. In conclusion, DFIA relaxes fault model requirements and more suitable than DFA when fault injection is hard to control.

\paragraph{\textbf{Safe Error Analysis (SEA)}} attacks exploit the dependence between the use of a faulty data and the value of a secret variable\cite{joye2002observability}. An adversary first identifies a target intermediate variable, of which use depends on the value of a secret variable. Then the adversary injects a specific fault into the target variable and observes whether the output is faulty or not. If the output is faulty, it means that the faulty target variable is used and the secret variable has a specific value. The advantage of the SEA is that it requires only a single-bit information from fault observation: If the faulty value has been used or not. Fault injection may be used to check if a specific computation is executed  (C-safe errors~\cite{yen2000checking}) or if a specific memory location is accessed (M-safe errors~\cite{karaklajic12}). SEA attacks have been successfully demonstrated on symmetric-key~\cite{blomer2003fault,joye12book} and public-key~\cite{yen2000checking,joye12book} algorithms. Yen and Joye describe a form of safe-error analysis that is based on collisions \cite{yen2000checking}. By forcing a value on an intermediate value with data dependency on the output, and by checking if the output is affected or not, a {\em collision} between the forced value and the original secret value can be detected. For this reason, for example, write-only cryptographic key registers should never allow partial update, otherwise the attacker can test a partial key guess by detecting these collisions.


\paragraph{\textbf{Algorithm-specific Fault Analysis}} uses fault injection to exploit algorithm-specific properties. For instance, the public-key cryptography algorithms such as RSA and ECC rely on a hard-to-solve mathematical problem. An adversary may use fault injection to alter the mathematical foundations of the problem and convert the problem into an easy-to-solve one.

In the infamous Bellcore Attack, Boneh et al. demonstrate that the security of the RSA cryptosystem can be broken with a single faulty computation~\cite{boneh1997importance}. In the RSA, a message $ M $ is signed by computing $ S = M^d\, modN$, where $d$ is the secret exponent and $ N = pq $ is a product of two large prime integers. The security of the system relies on the difficulty of factoring the modulus $N$. An efficient implementation of RSA is RSA-CRT, in which $S_1 = x^d\, modp$ and $S_2 = x^d\, modq$ are computed first and then the Chinese Remainder Theorem (CRT) is used to combine $S_1$ and $S_2$ to obtain $ S = M^d\, modN$. In the Bellcore Attack, a single random fault is assumed in the computation of either $S_1$ or $S_2$. If a fault occurs during the computation of $S_1$, the modulus $N$ can be easily factored using the equation $gcd(S-\hat{S}) = q$. In this equation, $S$ and $\hat{S}$ denote the fault-free and faulty signatures, respectively. Similar algorithm-specific analysis attacks have been mounted on several public-key systems including RSA and ECC~\cite{ciet2005elliptic,fouque2008fault,joye12book}.

\subsection{Using Fault Injection to Assist Side-Channel Analysis}
Another use of fault injection is to assist side-channel analysis for reducing the complexity of side-channel attacks or thwarting the countermeasures. Side-Channel Analysis, introduced by Kocher et al.~\cite{kocher1999differential}, is a major category of the implementation attacks used for cryptanalysis of secure embedded software. While fault attacks actively manipulate the physical operating conditions of a target device, side-channel attacks exploit physical leakage (e.g, power consumption, electro-magnetic radiation, etc.) emanating from the device during a security-critical operation. Side channel attacks are usually partitioned into two categories. \textit{Simple Side-Channel Analysis (SSCA)} exploits a single observation of the physical leakage of the device during a cryptographic operation. \textit{Differential Side-Channel Analysis (DSCA)} collects multiple observations of the physical leakage and retrieves the secret information by applying statistical tests on these observations. In the last 20 years, various SSCA and DSCA methods have been demonstrated on all forms of cryptography~\cite{fan2010state,oswald2013implementation,spreitzer2017systematic}. Similarly, developing countermeasures against SSCA and DSCA attacks have been extensively investigated~\cite{witteman2008,tillich2008attacking,rivain2010provably,grosso2014masking,chevallier2004low}. The advancements in the countermeasure design motivated \textit{fault-assisted side-channel attacks}, which utilize fault injection to break the security of systems protected against SCAs. 

In 2006, Skorobogatov~\cite{skorobogatov2006optically} used a laser source to illuminate a specific area of an SRAM memory to increase the side-channel leakage of the illuminated area. In 2007, Amiel et al.~\cite{amiel2007passive} proposed a fault-assisted side-channel attack on an RSA implementation resistant to both side-channel and fault attacks. In this attack, the injected fault modifies a secret variable such that the modified variable leaks information via SSCA. A similar attack was also developed by Clavier et al.~\cite{clavier2010passive} on an AES implementation protected with first-order masking, a DSCA countermeasure. Roche et al.~\cite{roche2011combined} also demonstrated a combined attack on an high-order masked and DFA-resistant AES implementation. Based on the work of Roche et al.~\cite{roche2011combined}, Dassance et al.~\cite{dassance2012combined} developed combined attacks on the key schedule of a protected AES implementation. In 2010, Schmidt et al.~\cite{schmidt2010combined} both demonstrated novel fault-assisted side-channel attacks and countermeasures on them. Later, Feix et al. proposed novel attacks that are capable of breaking the countermeasures proposed by Schmidt et al.~\cite{schmidt2010combined}. In 2018, Yao et al.~\cite{yao2018fault} proposed a fault-assisted side-channel attack that utilize fault injection to weaken a DSCA-resistant masking scheme and breaking its security with a first-order DSCA.  

\subsection{Fault-Enabled Logical Attacks}   

In addition to their use in cryptanalysis, fault attacks can also be used to trigger logical attacks (e.g, control flow hijacking, privilege escalation, subverting memory isolation, etc.) on smartcards and general-purpose processors. Classic logical attacks such as buffer overflow tamper with the inputs of a program to exploit a security bug in the implementation of the program. A well-known example is the HeartBleed bug \cite{bleed14}. In the absence of such an exploitable software bug, it is not possible for an adversary to mount a logical attack by just modifying inputs. In such a case, an adversary can inject faults to dynamically create required conditions to mount a logical attack. A straightforward application of this idea are attacks on input/output routines, which copy a portion of an internal memory region to the outputs of the chip. By glitching the end condition of the input/output routine, an adversary can force dumping of the entire internal data memory region, rather than just the portion allocated to the input/output buffer. Similarly, an attacker can also utilize fault injection to dump the source/binary code of the target program in case the code is normally not available to the attacker~\cite{obermaier2018shedding,micah}. This enables the attacker to analyze the source/binary code for identifying and subsequently exploiting the software vulnerabilities. The following paragraphs briefly explain fault-enabled logical attack examples from the literature.

Barbu et al.~\cite{barbu2010attacks} demonstrated two fault-enabled logical attacks on a Java card. In the demonstrated attacks, the adversary uses a laser-induced instruction-skip model to create type confusion. Then the adversary exploits the induced type confusion to load an unverified adversary-controlled code on the Java Card. Type confusion also enables an adversary to access other applications' memory space. Vetillard et al.~\cite{vetillard2010combined} and Bouffard et al.~\cite{bouffard2011combined} also demonstrates similar attacks on Java Card, in which they employed fault injection to bypass run-time security checks and execute malicious code on the platform.

The first fault-enabled fault attack on a general-purpose processor has been demonstrated by Govindavajhala et al.~\cite{govindavajhala2003using}. In the demonstrated attack, the adversary designs and runs a software program on a Java Virtual Machine (JVM) on a desktop computer. The malicious program is designed such that a bit error in the data space of the program allows the adversary take full control over JVM. To induce those exploitable faults, the authors overheat the memory chips. 

Nashimoto et al.~\cite{nashimoto2017buffer} proposed a fault-enabled buffer overflow (BOF) attack on a buffer overflow countermeasure, which limits input size. The authors demonstrated the proposed attack on an 8-bit AVR ATmega163 and a 32-bit ARM Cortex-M0+ microcontroller. Their fault models were single and multiple instruction-skip, which are induced by clock glitching. 

Timmers et al.~\cite{timmers16} demonstrated two ARM-specific, fault-enabled logical attacks which are based on setting the program counter (PC) of a microprocessor to an adversary-controlled value. The authors alter the execution of a memory-load instruction (i.e, instruction replacement) via voltage glitching to set PC to an adversary-controlled value. The authors provide two case studies to demonstrate the use of such an attack. In the first case, the authors bypass a secure-boot mechanism and run their own unverified program on the processor. In the second case, the authors subvert the hardware-enforced isolation mechanism of a Trusted Execution Environment (TEE) and run their code program with the highest privileges on the processor.  

Vasselle et al.~\cite{vasselle2017laser} demonstrated a fault-enabled logical attack on a Quad-core ARM Cortex-A9 processor, which bypasses secure boot mechanism and allows an adversary to get highest privileges on the processor. The authors achieved privilege escalation by resetting the privilege-level-specifying bit of the Secure Configuration Register of the processor via laser fault injection.

Timmers et al.~\cite{timmers17} proposed three fault-attack enabled logical attacks on a Linux Kernel to gain kernel-level execution privileges. The authors demonstrated their attacks on an ARM Cortex-A9 processor through voltage glitching. In the demonstrated attacks, the authors request system calls from the user space, and then, inject faults during the execution of system calls for privilege escalation. The gained privileges may allow an adversary to run an arbitrary code on the device and access the memory space of other applications.

The software-controlled fault injection methods such as triggering memory disturbance errors broaden the scope of fault attacks as they allow remote fault attacks. For example, in Rowhammer attacks~\cite{kim2014flipping}, an adversary-controlled program (running in a user space) injects bit-flip faults into security-sensitive DRAM memory cells by repeatedly accessing nearby cells. In 2015, Seaborn~\cite{seaborn2015exploiting} demonstrated two practical Rowhammer attacks. The first attack induces bit-flips to escape from Google Native Client (NaCl) sandbox. The second attack use bit-flips in DRAM for privilege escalation. Gruss et al.~\cite{gruss2016rowhammer} successfully mounted a Rowhammer attack from web browsers on four off-the-shelf laptops. Similarly, van der Veen et al~\cite{van2016drammer} achieved privilege escalation on Android-running mobile platforms. Razavi~\cite{razavi2016flip} demonstrated a Rowhammer attack in a cloud setting, in which a malicious virtual machine induce memory disturbance errors to gain unauthorized access to memory space of a co-hosted virtual machine. Kurmus et al.~\cite{kurmus17random} and Cai et al.~\cite{cai2017vulnerabilities} demonstrated that software-controlled memory disturbance errors can be triggered on Multi-cell (MLC) NAND Flash memories to mount fault-enabled logical attacks.

Finally, Tang et al.~\cite{tang2017clkscrew} exploited security-oblivious dynamic voltage and frequency scaling (DVFS) interface to induce faults in a smartphone. They demonstrated two software-controlled fault attacks. The first attack allows a malicious user-space program to inject faults into the operation of an encryption program running in Trustzone environment and to reveal the value of secret key stored in Trustzone environment. In the second attack, an adversary bypasses an authentication mechanism running in Trustzone to load an unauthorized program into Trustzone environment. These two attacks show that fault injection may enable an adversary to subvert hardware-enforced isolation mechanisms such as ARM Trustzone.

\subsection{Using Fault Injection to Assist Reverse Engineering}

Another potential use of fault injection is to assist reverse engineering. San Pedro et al.~\cite{san2011fire}, Le Bouder et al.~\cite{le2014fault}, and Clavier et al.~\cite{clavier2013reverse} employed fault injection to reverse engineer specifications of block ciphers similar to Data Encryption Standard (DES) and Adavanced Encryption Standard algorithms. For instance,  San Pedro et al.~\cite{san2011fire} propose the FIRE attack that employs fault injection to reverse engineer SBox specification of DES-like and AES-like block ciphers. In the AES version of the FIRE, single-byte faults are injected into the penultimate round of AES and faulty output data are collected. Then, the faulty data are converted into a set of linear Boolean equations. Finally, the equation system is solved using the Gaussian elimination and the SBox is reverse-engineered. Similarly, Jacob et al.~\cite{jacob2002attacking} induce faults into the execution of an obfuscated cipher and retrieve the secret key. Courbon et al.~\cite{courbon2015combining} also demonstrated a method to reverse-engineer gate-level structure of a hardware implementation of Advanced Encryption Standard (AES) algorithm, in which laser fault injection and image processing are combined.
\section{Fault Attack Evaluation and Certification}
\label{sec:evaluation}
Given the abundance and diversity of attacks on hardware products, the question arises for individual products: What is the attack resistance, and how secure does the product need to be?  
This question has been addressed by the global security certification community, resulting in the Common Criteria (CC)~\cite{commoncriteria1}. This standard defines levels of security and a methodology for evaluation. A consortium of vendors, certification bodies, and labs maintain a procedure for attack rating.

Any product with secure hardware, for which a vendor seeks certification, can be evaluated according to this methodology. The aim is to provide sufficient assurance that the product remains secure during several years of operation. In this section, we first briefly explain the steps and actors involved in the certification process. We then demonstrate how the fault attack resistance of a product is evaluated.  

\subsection{The Common Criteria Certification Process}
Three main roles exist in the certification process:
\begin{itemize}
\item \textbf{Vendor} is the manufacturer that develops the product to be certified.
\item \textbf{Evaluator} is the security evaluation lab that reviews and tests the
product design and implementation.
\item \textbf{Certification body} is the authority that certifies the product after successful completion of the evaluation.
\end{itemize}
Additional roles may be involved for separate software/hardware vendors, for product issuers, and evaluation sponsors, but these roles do not significantly change the basic model.
\begin{figure}[t]
	\centering
	\includegraphics[width=0.8\columnwidth]{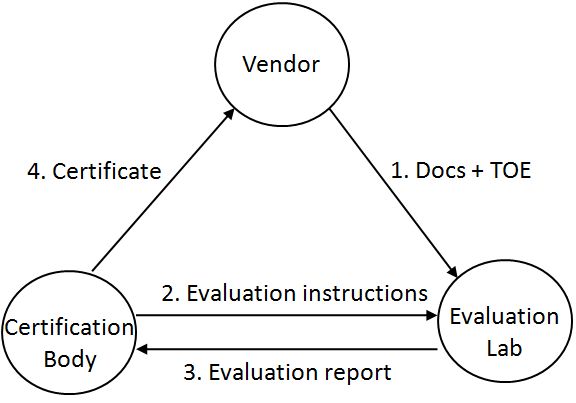}
	\caption{The Common Criteria (CC) Certification Process.}
	\label{fig:cc-process}       
\end{figure}

Figure~\ref{fig:cc-process} depicts a simplified view of how the three main actors work together. The process takes the following steps:
\begin{enumerate}
\item A vendor who wants to have a product CC certified provides an evaluator with all relevant documents and the product to be certified (i.e, Target of Evaluation--TOE).Vendors need to create and maintain a significant set of documents describing the system design and implementation in detail. These documents should also include proof that the implementation meets its specifications. The first task for the evaluator is to review all documents and decide whether they conform to the CC methodology and its requirements.

\item The certification body provides instructions to the evaluator as to how to evaluate the product. The evaluator performs all relevant tests to prove resistance of the product and informs the certification body. Strictly, the evaluator verifies that the product conforms to its security target, but the scope of that must be approved by the certification body. If any blocking issues are found, the product may need revision and re-evaluation.

\item The evaluator sends the Evaluation Technical Report (ETR) to the certification body (and to the vendor) who then reviews the evidence reported by the evaluator. Before issuing a certificate, the certification body may mandate additional testing if new threats surfaced, or when the test results gave reason to doubt. 

\item If no objections remain, the certification body issues the certificate.
\end{enumerate}
A product typically consist of multiple layers (hardware, operating system, applications) which can be independently certified. This would start with the hardware certification, after which composite evaluations can be done, where a new layer is certified in conjunction with a certified platform.

The CC certification process is known to be cumbersome; it is lengthy and costly~\cite{gao2006}. Although completion of the process may take a couple of months in an ideal situation, it often takes much longer. Apart from cost and time-to-market, this also carries a security penalty: Vendors may be hesitant to make security improvements to a certified system since changes break the certification. In this way, vulnerabilities may remain longer in products than needed. 

While Common Criteria certifications enjoy popularity for high assurance products, it is not the only evaluation methodology. Other schemes exists such as EMVCo~\cite{emvco} and FIPS 140-2~\cite{fips-140-2}, many of whom are lighter in execution. However, the processes described here are similar across many schemes and serve well to explain the ecosystem. Next, we will take a closer look into the evaluation process.

\subsection{Evaluation of Fault Attack Resistance}
Products are always evaluated in ‘white-box’ style. This means an evaluation lab gets access to all product design and implementation information. This will reduce evaluation cost (no need for lengthy reverse engineering), and reduce the risk of missing big security weaknesses.
Typically an evaluation consists of two phases:
\begin{enumerate}
\item Vulnerability Analysis (VA)
\item Penetration Testing (PT)
\end{enumerate}
During the VA, the lab reviews the design and implementation code of a product, and weighs this against applicable threats.
In the PT, a product is tested against a number of attacks to measure its actual resistance. All successful attacks are rated, and when their scores are sufficiently high, the product qualifies for certification.

The evaluation methodology distinguishes between Identification and Exploitation. The former defines the cost of demonstrating that the attack works on the product, while the latter looks at the cost of repeating the attack. Both aspects are important. For instance, an attack with very high initial cost will scare away low-budget attackers, while a high repetitive cost will prohibit attack scaling.

Parameters used in the attack rating are: 1) time, 2) expertise, 3) product knowledge, 4) number of the target samples, 5) equipment, 6) ability to configure target.
The parameter time is extremely important during an evaluation as this is most often the limiting aspect during the penetration testing phase. An evaluator cannot afford to lose time if several attacks are to be executed within the typical time frame of a few months. 

\begin{figure}[t]
	\centering
	\includegraphics[width=1\columnwidth]{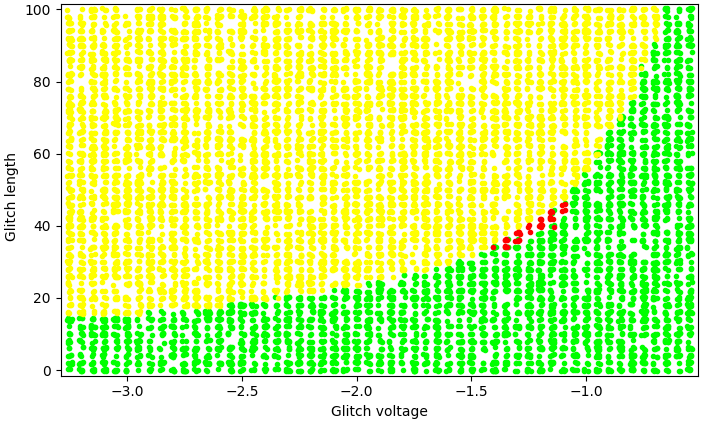}
	\caption{Relation between the glitch length, glitch voltage, and fault behavior.}
	\label{fig:fi-pic3}       
\end{figure}

\begin{figure}[t]
	\centering
	\includegraphics[width=1\columnwidth]{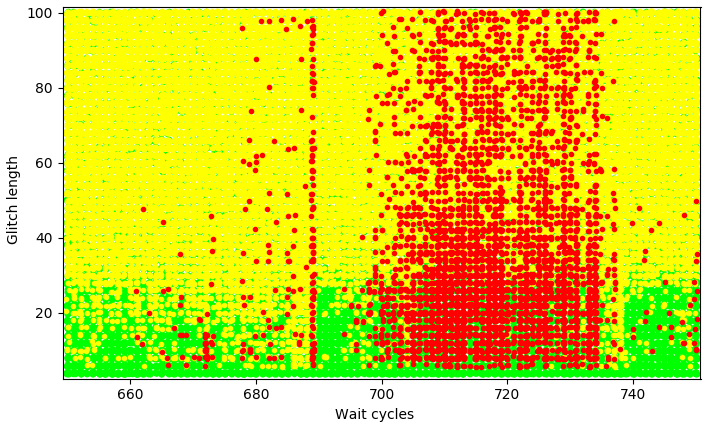}
	\caption{Relation between the glitch length, temporal offset (i.e, wait cycles), and fault behavior.}
	\label{fig:fi-pic4}       
\end{figure}

For efficiency, the PT starts with an investigation of sensitivity to different fault injection methods. While Voltage glitching is typically the simplest method, this is often prevented by sensors. Alternative methods like EM and optical glitching are more complex, but also harder to prevent. During this sensitivity analysis the evaluator uses test software on the target that runs loops and typical instructions that may be affected. The test software accelerates the detection of hardware weaknesses, and supports finding optimal attack parameters, such as glitch intensity and duration. Figure~\ref{fig:fi-pic3} shows how the right combination of glitch voltage and glitch length can be found. The green dots represent experiments that did not affect the chip. Alternatively, the yellow dots represent experiments where the glitch was too strong and resulted in a reset of the chip. Finally, the red dots represent successful glitch parameters that resulted in a an observable effect in the test code. Figure~\ref{fig:fi-pic4} shows how effective temporal offsets are found (represented as wait cycles on the x-axis).

Once a weakness is established, a setup is built to demonstrate that the weakness can be exploited on the real product software. This setup includes equipment for generating glitches at the right time, and solutions to minimize the effect of countermeasures. A highly automated setup runs for several days and uses dedicated control procedures to manage smooth repetition and fault logging.
Ultimately, an evaluation results in a report, where successful attacks are rated, and a discussion is given on the attack risks and potential mitigation strategies.

\section{Conclusions}
\label{sec:conclusion}
This paper provides a review on mechanisms, implementation, and evaluation of hardware-based fault attacks that aims at breaking the security of embedded software. Our review shows that multiple abstraction layers (software, instruction-set, and hardware layers) take part in fault attacks and each abstraction layer may be a target for an adversary. We also observe that fault attacks breaks a fundamental assumption made by secure embedded software: The hardware layer of the embedded systems ensure the correct execution of the embedded software. Therefore, fault attacks pose a serious security threat to any kind of security-critical software (firmware, operating system, user applications, and cryptography) running on embedded devices:
\begin{itemize}
\item In a fault attack on embedded software, the target of fault injection is the hardware layer while the target of exploitation is the software layer. Thus, it is not always possible to mitigate the fault attack threat with software-only countermeasures.

\item Fault attacks do not require presence of a software bug in the embedded software because the fault attacks dynamically alter the behavior of the underlying hardware through fault injection. As our review shows, this enables fault attacks to dynamically induce software bugs, to thwart countermeasures, and to enable other attacks.

\item Although fault attacks traditionally require expertise and expensive equipment, they tend to become more accessible because of the advancement in the field. As we demonstrate, today, it is possible to inject faults via inexpensive hardware equipment (less than 500\$) or via only software programs.   
\end{itemize}
Considering the increasing role of the embedded Internet of Things (IoT) devices in our daily life and critical infrastructure, we believe that embedded hardware and software developers need to put additional effort on mitigation and evaluation of the fault attack risk on the embedded systems.   

\section*{Acknowledgements}
The authors would like to thank Dennis Vermoen from Riscure Security Lab for his help and support. During this work, the first author was supported in part through the National Science Foundation Grant 1441710 and 1314598, and in part through the Semiconductor Research Corporation.

\bibliographystyle{spbasic}      
\bibliography{hass2018}

\end{document}